\documentclass[12pt,twoside]{article} 
\usepackage{graphicx}

\begin{document} 


\centerline{} 

\centerline{} 

\centerline {\Large{\bf Non-relativistic model for interaction}} 

\centerline{} 

\centerline{\Large{\bf of dark particles with matter}} 

\centerline{} 

\centerline{\bf {N. Takibayev}} 

\centerline{} 

\centerline{Institute of Experimental and Theoretical Physics} 

\centerline{al-Farabi Kazakh National University} 

\centerline{al-Farabi 71, Almaty 050040, Kazakhstan} 



\begin{abstract} A non-relativistic quantum mechanical model for interaction of dark  particles with matter is proposed; the model describes resonant amplification of effective interaction between two massive bodies at large distances between them. The phenomenon is explained by catalytic action by a third dark particle scattered at a system of two heavy bodies. It is shown that effective interaction between the heavy bodies is changed at larger distances and can transform into repulsion contributing in that case to the dark energy action.
\end{abstract} 



\section{Introduction} 
As it is known, the dark matter problem arose from analysis of astronomical data when considerable discrepancy between orbital velocities of stellar objects at periphery of large galaxies compared to velocities calculated for usual gravitational fields has been found (see, for instance, reviews in [8, 14]). This phenomenon was explained by the dark matter action that resulted in increased mass of such stellar objects. Discovered effect of gravitational lensing considerably supported such assumption.

A particular obstacle was to provide experimental verification for existence of the dark matter particles. Many candidates for such particles have been proposed as well as several theories for their interaction with ordinary matter were considered [1, 15]. Still, particles of dark matter have not been found. Another problem was caused by relatively low density of dark matter inside galaxies, where the theories [14, 15] predicted formation of dark matter clouds.

Present paper proposes a model for resonance amplification of interaction between two astronomical stellar objects stimulated by additional interaction with dark matter particles (hereinafter dark particles). In other words, it is considered that the dark  particles not just increase gravitational mass of stellar objects, but, more importantly, do noticeably change effective interaction between the two heavy bodies. Such changes became distinguishable at specific large distances between the heavy particles only.

We refer to the two heavy stellar objects as to a system of one massive body located in a center of a galaxy and another stellar body located at the galaxy$'$s periphery orbiting the center of the galaxy.

Resonant amplification of the effective interaction between the two heavy bodies imitates increase in their mass while their true gravitational mass does not change at that. Such increased interaction results in more pronounced gravitational lensing of the bypassing light. So, the model can describe main effects and phenomena of the dark matter problem.

For simplicity, we consider the task in the non-relativistic case. Our considerations below are presented within the quantum-mechanical theory of the three-body scattering. Scattering of a dark particle at the system of two heavy bodies is considered. The heavy bodies in our considerations are massive stellar objects as described above and a dark particles has a negligibly small mass, for example, in compared with the electron mass.

We propose a simple potential model for interaction of a dark particle with a massive body; such approach allows solving the three-body problem analytically. These solutions distinctly demonstrate the resonance dependence of the effective interaction between two massive bodies on distance between these bodies.

So, at relatively small distances the catalytic amplification is negligibly small and effective interaction here coincides with direct (gravitational) interaction between the heavy bodies. At higher distances, catalytic action of the dark particles increases, and effective interaction of the heavy bodies also increases. At even larger distances, catalytic action can reverse and, instead of attraction between the heavy bodies, cause their repulsion. This to certain degree can imitate dark energy action.

In reality, the nature of interactions between dark particles and a heavy body can be more complex and the catalytic action may happen to be far from simple.

\section{Model quantum-mechanical scattering problem for a three-body system} 
Let us assume that the dark particles do not interact with each other. This simplifies the problem. Of course, we must take into account that it is necessary to consider the flux of particles per unit time, the total action of which is accepted here as the action of a single particle, taken with a certain energy and momentum.
We would also not consider the three-body type forces.

Solutions for scattering amplitudes in the three-body problem are based on solutions for scattering amplitudes in two-body subsystems [5,6]. For simplicity, let us set the pair interaction of a dark particle with a heavy body by a potential in separable form. 

The task therefore is reduced to determination of interaction in a system of three bodies: two heavy ones interacting gravitationally and one dark particle interacting with each of the heavy bodies.  

Another simplification will be related to utilization of the Born-Oppenheimer approximation which leads to the three-body problem solution in analytical form [16]. Such approximation can be used here since the mass ratio of the dark particle and the heavy body is extremely small.

Let us take the potential of two heavy body interaction in non-relativistic limit and take its dependence on distance same as of coulomb interaction, for simplicity. In non-relativistic quantum mechanics the Hamiltonian $H$ of heavy body with mass $M$ in Newtonian gravitational field is given by:
\begin{equation}
\hat{H} = - \frac{\hbar^2}{2M} \nabla^2 + M\Phi_G (\vec{r}) \  \  ,
\end{equation}			
where $\Phi_G (\vec{r})$  is Newtonian gravitational potential at distance $r$. With no magnetic fields presented, this equation is the same as Schr$ö$dinger equation and the gravitational potential $\Phi_G (\vec{r})$ is the same as for the classical field [7]. The equation (1) can be generalized by inclusion of additional non-Newtonian forces (see, for example in [3, 7]). 
In the present paper we choose the simplest interaction to clearly reveal the resonant amplification of the effective interaction between the heavy bodies.

The Faddeev equations [5, 6] are the key to solve the three-body scattering problem. When a system of two heavy bodies and one dark particle is considered, these equations in combination with the Born-Oppenheimer approximation lead to simplification of the initial equations. At that, for a set of model pair potentials the solutions can be obtained analytically what makes the math better comprehended.

Additional amplification of interaction between the heavy bodies due to their interaction with the dark particle is determined. This third particle acts as a catalyst particle (some comments in [10]). Unlike regular pair interaction between heavy bodies, the additional  effective interaction can resonantly increase at certain distances between the heavy particles at certain impulse of the dark particle. Let us call such certain distances the resonant ones.

The amplification or additional forces decrease abruptly at distances considerably less or, on the contrary, larger than the resonant distance. As a result, regular spatially monotonous pair interaction forces become dominant again [16, 17].

\section{A model for potential between a dark particle and a heavy body}

Let us use a system of units where $c=1$, $\hbar=1$. We also assume that energy of a dark particle is small, i.e. we deal with non-relativistic particles. Interaction of the dark particle and the heavy body in separable form can be described as:
\begin{equation}
V_{DH}=|\nu>\lambda_{DH}<\nu|  \; .
\end{equation}
Here $\lambda_{DH}$ is the coupling constant, the $DH$ subscript denotes interaction of the $Dark$ particle and the $Heavy$ body. Then, based on the Lippmann-Schwinger equation 
\begin{equation}
T_{DH}=V_{DH} + V_{DH} G_0 (E) T_{DH}  \; ,
\end{equation}
where $G_0 (E)$ - is the free Green$'$s function of system; one can get the solution for the $T_{DH}$-matrix also in separable form:
\begin{equation}
T_{DH}=|\nu>\eta_{DH} (E) <\nu|  \; , 
\end{equation}
\begin{equation}
\eta^{-1}_{DH} (E) = 1/\lambda_{DH} + I(E)  \; , 
\end{equation}
\begin{equation}
I(E) = - \int d\vec{p}  \frac{\nu^2 (p)}{E-E_s+i\gamma}  \; . 
\end{equation}
$E = p^2_0/2\mu$, $E_s = p^2/2\mu$, $\mu$ - denotes reduced mass of the pair. Hereinafter we use the following normalization: $d \vec{r} = r^2 dr d(\cos(\theta) d\varphi  $, but $d \vec{p} = p^2 dp d(\cos(\theta) d\varphi  /(2\pi)^3$.

For simplicity, let us consider the $S$-wave and the potential form-factor in the form:  
\begin{equation}
\nu (p) = N_{DH}\frac{t}{1+t^2} 
\end{equation}
where $t=p/\beta$, $N^2_{DH}= 4\pi/(\mu\beta)$; we get ($t_0=p_0/\beta$, $\beta$ - – potential parameter)  
\begin{equation}
I(E) = I(t_0)= \frac{1-2it_0}{(1-it_0)^2}  \; . 
\end{equation}
	
 We have to normalize  $I(t_0 = 0) = 1$ to make the coupling constant to be dimensionless. Then, $\lambda_{DH} < 0$ would correspond to the attraction forces. For $ -1 < \lambda_{DH} < 0 $, the amplitude pole corresponds to the virtual level, and for $\lambda_{DH} < -1$ it corresponds to the bound state since for $t_0 = i\tau$,  
$\tau = - (1 + \lambda_{DH})$. Here and below $\lambda_{DH}$ is real by definition.  

The value $\eta_{DH}$ is frequently referred to as an amplification coefficient to highlight the similarity between the potential and the $T_{DH}$-matrix forms as well as the dependence of $\eta_{DH}$ on the initial energy of the pair.

In the case of $\lambda_{DH} > - 1$ we would have no bound states in the system of a dark particle and a heavy body. Note, in the case of $-1 < \lambda_{DH} < - 1/2$ the solutions for the two-body amplitude can be presented in Breit-Wigner form.

Our goal is to determine the behavior of the effective interaction between the heavy massive bodies as a function of the distance between them.

\section{Scattering of a dark particle on a pair of heavy bodies}

Mathematically rigorous solution of the three-body problems was provided by L.D. Faddeev [5]. Faddeev equations for components of the three-particle $T$-matrix can be written as [6]:
\begin{equation}
T_{ij} (E) = T_i \delta_{ij} + T_i G_0 (E) \sum_{l\neq i} T_{lj} (E) , \; \; i,j = 1,2,3
\end{equation}
where $T_i = V_i + V_i G_0 (E) T_i $, $V_i$ are pair interaction potentials and $T_i$ - their corresponding pair $T$ -matrixes, $\delta_{ij}$ - Kronecker delta. Usually for conciseness, the interacting pair of particles is denoted by the number of the third particle. The complete $T$-matrix would correspond to the sum: $T = \sum T_{ij}$. The index $i$ in $T_{ij}$ stands for the number of the pair which interacts last at the left asymptotic, i.e. corresponds to the number of a particle which first escapes the interaction region. Similarly, the index $j$ denotes the number of the pair which interacts last at the right asymptotic. 

Let us first solve the problem taking into account interaction of a dark particle with the heavy body: at this stage we assume that there is no pair interaction between the two heavy bodies. The pair of heavy particles is denoted with $1$. Two other subsystems with one dark and one heavy body are denoted by $i,j = 2,3$, i.e. $i,j \neq 1$. Their potentials are taken in the separable form (2).  

Related components $T_{ij}$  with non-connected and connected parts can be written as [16]: 
\begin{equation}
T_{ij} = T_i \delta_{ij} + |\nu_i>\eta_i M_{ij}\eta_j <\nu_j|  \; ,
\end{equation}
and transferred into the matrix equations for $M_{ij}$  
\begin{equation}
M_{ij} = \Lambda_{ij}  + \sum_{l\neq 1} \Lambda_{il}\eta_l M_{lj}   \; ,
\end{equation}
where $\Lambda_{ij} = <\nu_i| G_0 |\nu_j>$, $j \neq i$.
It is important that the diagonal elements of this matrix are identically equal to zero, i.e. $\Lambda_{ii} \equiv 0$. This peculiarity of the Faddeev equations assures compactness of the core of the integral equations [5,6,9].

Let us consider a limiting case when $m/M \rightarrow 0$, where $m$  - mass of a dark particle and $M$ - mass of a heavy body.  
In this limit the total energy of the system is $E = p^2_{01}/m $, where $\vec{p}_{01} = \vec{p}_{0}$	- initial momentum of the
dark particle. Pair interactions between a dark particle and heavy bodies would not depend on momentums of the heavy particles $<\nu_2|\vec{p}> \rightarrow \nu(\vec{p})$,  $<\nu_3|\vec{p}> \rightarrow \nu(\vec{p})$, where $\vec{p}_1 = \vec{p}$. Correspondingly, $\eta_2 = \eta_3 \rightarrow \eta(p_0) $  would also be the functions of initial momentum of the dark particle.  

  The value  $\Lambda_{ij}$  takes the form:
\begin{equation}
\Lambda_{ij} = 2m\frac{\nu_i(\vec{p}) \nu_j(\vec{p})}{p^2_0 - p^2 + i\gamma} = 
f(\vec{p}, p_0) \;  \;  , \;  \;  i\neq j \; ,
\end{equation}
here $\vec{p} = \vec{p}_1 = -\vec{p}_i - \vec{p}'_j$. We denote with the prime the variables of heavy particles at the reaction outlet; those without a prime $–$ at the reaction entrance. $\Lambda_{ij}$  can be written in the form: 
\begin{equation}
\Lambda_{ij} (\vec{p}_i,\vec{p}'_j) = \int d\vec{r} \exp(i\vec{r}\vec{p}_i) J_{ij}(\vec{r}) 
\exp(i\vec{r}\vec{p}'_j) \; \; ,
\end{equation}
where 
\begin{equation}
J_{ij}(\vec{r}) = \int d\vec{p} \exp(i\vec{r}\vec{p}) f(\vec{p};p_0) \; \; .
\end{equation}
Similar transformation for the amplitude gives us:  
\begin{equation}
M_{ij} (\vec{r}_i, \vec{r}'_j) = \int d\vec{r}'' \delta (\vec{r}_i - \vec{r}'') 
[J_{ij}(\vec{r}'') \delta (\vec{r}'' + \vec{r}'_j)    
+ \sum_{l=2,3}J_{il}(\vec{r}'') \eta_l (p_0) M_{lj} (-\vec{r}'', \vec{r}'_j)] 
\end{equation}

Since $\delta$-functions call off the integration in the right part of (15), the equation for 
$M_{ij} (\vec{r}, \vec{r}')$ takes a simple form:
\begin{equation}
M_{ij} (\vec{r}, \vec{r}') =  J_{ij}(\vec{r}) \delta (\vec{r} + \vec{r}') 
+ \sum_{l=2,3} J_{il}(\vec{r}) \eta_l (p_0) M_{lj} (-\vec{r}, \vec{r}') \; \; .
\end{equation}
Moreover, $\delta$-functions in (15) give the equalities: $\vec{r}_i + \vec{r}'_j = 0$ if $i \neq j$, and $\vec{r}_i - \vec{r}'_i = 0$. It means that $\vec{r}_i$ and $\vec{r}'_j$ are counted from the center of symmetry between the positions of the heavy bodies. In this case for $\vec{r}$ and $\vec{r}'$ the indeces $i$ and $j$ can be omitted.  
 
The solutions can be written in a convenient form [16]: 
\begin{equation}
M(\vec{r}, \vec{r}') =  M^+(\vec{r}) \delta (\vec{r} + \vec{r}') +  M^-(\vec{r}) \delta (\vec{r} - \vec{r}') \; \;  ,
\end{equation}
where
\begin{equation}
M^+(\vec{r}) = \sum_{l} [I - K(\vec{r}) \eta (p_0)]^{-1}_{il} J_{lj}(\vec{r})  \; \;  , \; \; j\neq i  \; \;  ,
\end{equation}
\begin{equation}
M^-(\vec{r}) = \sum_{l} [I - K(\vec{r}) \eta (p_0)]^{-1}_{il} K_{li} \; \;  ,
\end{equation}
and
\begin{equation}
[K(\vec{r})]_{ij} = K_{ij}(\vec{r}) = \sum_{l} J_{il}(\vec{r})\eta_l(p_0)J_{lj}(-\vec{r})  \; \;  .
\end{equation}

Here we note that in our three-body problem we deal with two variables: $p_0 \; - $ initial momentum (wavenumber) of the dark particle  and $r = d/2$  ($d \; - $ distance between the heavy bodies).

\section{Oscillating interaction between two heavy bodies}

In this section we denote with index $\phi$ the solutions and interactions in the three body system, but with the $V_1$ potential excluded. We also introduce notations $V_{\phi} = \sum V_i$, where $i = 2, 3$, and $G_{\phi}$  - Green$'$s function, and $|\Phi >$  - wave function for this system. The solutions are determined in (17) - (20). 

Now one can move forward to solve a problem with inherent pair interaction between heavy particles taking into account the potential  $V_1$. The pair potential $V_1$ acts directly between the two heavy bodies of the system. The Lippmann-Schwinger equation for scattering at two potentials can be written as follows [9, 10]:
\begin{equation}
|\Psi> = |\Phi> + G_{\phi} V_1 |\Psi>  \; \;  ,
\end{equation}
where $\Psi(\vec{r})$  - total wave function for a system of three particles.
This symbolic equation is a short form of the system of equations in matrix form similar to the Faddeev equations. The first term in (21) at right discribes the resonant behaviour, but without direct interactions between two heavy bodies. 

 Since
\begin{equation}
G_{\phi} = G_0 + G_0 T_{\phi} G_0   \; \;  ,
\end{equation}
where $T_{\phi}$  is the matrix with the components as in (10), we can rewrite (21) in a symbolic form  
\begin{equation}
|\Psi> = |\Phi> + G_0 (I + T_{\phi} G_0) V_1 |\Psi> = |\Phi> + G_0 V_{ef} |\Psi> \; \;  ,
\end{equation}
where 
\begin{equation}
V_{ef} = (I + T_{\phi} G_0) V_1    \; \;  .
\end{equation}
This effective potential corresponds to the distorted potential $V_1$
\begin{equation}
V_{ef} =  V_1 + \sum_{i,j = 2,3} |\nu_i>\eta_i (\delta_{ij} + M_{ij} \eta_j) 
<\nu_j|)G_0]V_1    \; \;  .
\end{equation}
It then follows that the effective potential as long as the amplitude $M_{ij}(r)$ would depend resonantly on both distance between the heavy particles and energy of the dark particle.

In order to compare the interaction potentials $V_1$ and $V_{ef}$, acting between the two heavy bodies, it is necessary to record the first potential in the three particle representation. For this purpose, $V_1$ can be supplemented with Born interaction of the dark particle. A similar addition should be made for the second potential - $V_{ef}$. Then the ratio of these quantities, written in the form
\begin{equation}
\xi  = \frac{<\Psi_0|V_i G_0 V_{ef} G_0 V_i |\Psi_0>}{<\Psi_0|V_i G_0 V_1 G_0 V_i |\Psi_0>}  = \frac{<\nu_i| G_0 V_{ef} G_0 |\nu_i>}{<\nu_i| G_0 V_1 G_0 |\nu_i>} \; \;  ,
\end{equation}
will lead us to the result of the interaction between heavy bodies due to participation of an additional dark particle.

Accordingly, the enhancement factor will be equal to:
\begin{equation}
\xi  = 1 + I(t_0) \eta_i (\delta_{ij} + M_{ij} \eta_j)     \; \;  .
\end{equation}
This important outcome demonstrates that the interactions with the third particle can substantially change the character of forces acting between the heavy bodies. 

Figures 1 and 2 show the behavior of real and imaginary parts of the enhancement factor $\xi = \xi(t_0,\rho)$ for $\lambda_{DH} = - 0.95$.  

\begin{figure}
\centering
  \includegraphics[natwidth=400,natheight=120]{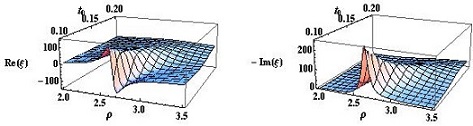}
\caption{Real and imaginary parts of $\xi(t_0,\rho)$; $\lambda_{DH} = -0.95$; the resonance at $\rho \approx 2.65$.}
\label{fig:1}       
\end{figure}

\begin{figure}
\centering
  \includegraphics[natwidth=350,natheight=150]{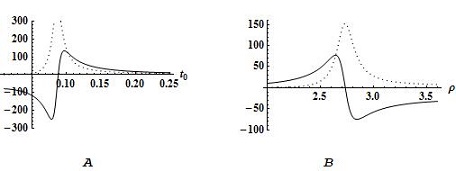}
\caption{The solid line shows the real part of $\xi$, while the dotted line shows the imaginary part of ($-\xi$): A) as function of $t_0$ at fixed value of $\rho = 2.6 $; 
B) as function $\rho $ for a fixed value of $t_0 = 0.12$; the resonance at $\rho \approx 2.65$; $\lambda_{DH} = -0.95$.}
\label{fig:2}       
\end{figure}

 Following the model in Section 4 above, one can get for $J_{ij}(\vec{r}) $: 
\begin{equation}
J_{ij}(\vec{\rho}) = \frac{\exp(-\rho)(1-2/\rho)}{1+t^2_0} - \frac{2}{\rho} 
\frac{\exp(-\rho) + t^2_0 \exp(it_0\rho)}{(1+t^2_0)^2}  \; \;  ,
\end{equation}
where dimensionless variables $ \vec{\rho} = \vec{r} \beta $ and $ t_0 $  are introduced. 

It is necessary to note that location of the amplitude pole $M^{\pm}_{ij}(\vec{\rho})$  on the $\rho, t_0$  plane would change depending on the coupling constant $\lambda_{DH}$ (Fig. 3). 
But the key factor here is the resonant dependence of the amplitudes $M^{\pm}_{ij}(\vec{\rho})$  on distance between the heavy particles.

It is important that the resonances take place not only when distance between the heavy bodies varies, but when the dark particle wavenumber changes $p_0 = t_0 \beta$. It means that even at resonant distances the resonance takes place only at certain wavenumber values of the dark particles. This is a distinguishing feature of resonances in quantum-mechanical few-body systems as regards the continuum.

Such phenomenal behavior of the effective interaction between the heavy particles is due to three-particle quantum-mechanical effects.

Here one should note that the described above model can potentially help explaning  phenomenological interactions at galactic scales.

\begin{figure}
\centering
  \includegraphics[natwidth=400,natheight=120]{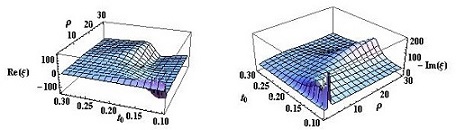}
\caption{Real and imaginary parts of $\xi(t_0,\rho)$; $\lambda_{DH} = -0.97$; resonance regions at $\rho \approx 3$ and $\rho \approx 20$.}
\label{fig:3}      
\end{figure}

Let us assess the characteristic distances. For example in our model, resonance takes place at the following parameters: $\lambda_{DH} = - 0.95$, $\rho = r \cdot \beta = 2.5$. 
Assuming that the distance between the massive bodies is about $r \approx 2.5 \cdot 10^{22} cm$, one can get $\beta \approx 10^{-22} cm^{-1}$ and $p_0 \geq 10^{-23} cm^{-1}$.

Then, taking into account the resonant behavior of the effective interaction at such distances (see Fig. 2), one can get from (27) the following multiplication of the interaction potential for the factor: $\xi \approx 100$ at $\rho = 2.5$. 

So we can say that exchange dark particle flux between heavy gravitating bodies acts like the dark matter. The orbital velocity of the peripheral body becomes higher than that at normal gravity owing to  the enhancement factor. Moreover this flux would glue up these two heavy bodies and, with respect to other particles and fields (gamma quanta, for instance) such a system would appear as a single object. The system would have the effective mass much higher than its own mass.

It is quite possible that such mechanism can contribute to gravitational lensing of electromagnetic radiation.

It is remarkable that the enhancement factors can be negative at certain values ρ in the case of two-body attractive potentials (see Figs 1$-$4). For example, in the case of $\lambda_{DH} = -0.95$ we have $\xi \approx - 80 - i\cdot 50$ at $\rho = 2.85$ . Then the resulting potential between two heavy bodies looks like a repulsive force.  Here the question arises - to what extent can this potential also simulate the action of dark energy? 

Fig. 4 shows the dependence of $\Xi(\rho)$ on $\rho$ at $\lambda_{DH} = -0.95$. 
$\Xi(\rho)$ is the overall result of $\xi(t_0,\rho)$ on the interval $0.001 < t_0 < 0.6$.

The opposite example is given in Fig. 5 that demonstrates $\xi = \xi(t_0,\rho)$ values in the case of repulsive potential between the dark particle and heavy body 
($\lambda_{DH} = 10$). 

\section{Discussion}
\label{discussion}

In the proposed paper we consider a region of the continuous spectrum (continuum). Coupling constant is taken as $\lambda_{DH} > -1$, so the bound states in the pair of the dark particle and the heavy body does not arise. Moreover, we consider the reverse limit when the interaction radius of $DH-$forces is taken large, but its ratio to the scattering length remains unchanged 
$\sim (1 + 1/\lambda_{DH})^{-1}$.

The direct gravitational interaction between two heavy bodies in the model is complemented by interaction with the third dark particle (or a flow of dark particles in order to be more exact). This additional interaction has a specific resonant behavior as a function of the distance between the heavy bodies. The resonant behavior of the effective potential implies its significant increase at certain distances between these bodies.

\begin{figure}
\centering
  \includegraphics[natwidth=250,natheight=120]{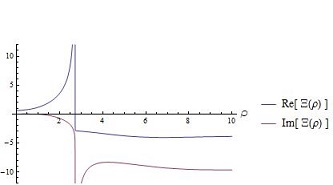}
\caption{Real and imaginary parts of $\Xi(\rho)$; $\lambda_{DH} = - 0.95$.}
\end{figure}

The figures 1$-$4 show that the strengthening of effective interaction occurs at certain distances between the heavy particles while at other distances we observe sharp decrease or even repulsion. This particular fact reveals a certain similarity in the action of such effective potential and dark matter.

As it follows from the analysis of astronomical observations, dark matter in dwarf galaxies appears to be quite small [12, 13], but its effect is huge in big galaxies [11, 14]. I.e. dark matter is stronger between massive bodies located in the center and on the periphery of a galaxy in large enough galaxies.

Here, we assume that the ensemble of heavy particles in a distant star should be considered as a single massive body, as well as an ensemble of heavy particles in another massive body located in the center of the galaxy. These massive bodies will interact with each other not only via gravity but also through exchange flows of dark particles. 

There is another important fact. Ultra-cold dark particles, apparently cannot be registered in terrestrial environments. Indeed, the corresponding wavelengths are large or even of galactic dimensions. These dark particles do not form bound states with other particles and massive objects. Their impact will be insignificant within the Earth and the Solar System [13]. 

However, the effect of these particles within large galaxies may already be quite noticeable. Strengthening of effective attractive forces will alter the orbits of massive objects in such galaxies.

\begin{figure}
\centering
  \includegraphics[natwidth=400,natheight=120]{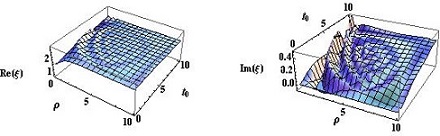}
\caption{Real and imaginary parts of $\xi(t_0,\rho)$; $\lambda_{DH} = 10$.}
\end{figure}

At even greater distances than the galactic ones, this model provides effective repulsive forces. It requires further investigation to understand how this effect simulates the dark energy action.

Of course, this model of pair interaction is very simple and can be improved. For example, one can also include in a consideration $P$-wave and high waves, and several kinds of dark particles. Accordingly, the sum interactions will be more complex.

\section{Conclusion}
\label{conclusion}

The effect obtained within the proposed model imitates the dark matter amplification of interaction forces at very large but definite distances.

The important feature of the resonant amplification is its selectiveness. The effect is negligible at distances much less or, vice versa, much greater than the resonance distances. This resonant amplification is based on well-known quantum-mechanical considerations and, in particularly, on the quantum considerations of few-body interactions. 

We hope that application of these theories and methods in physics of super-large distances will also be fruitful and exciting. 





\begin{thebibliography}{99} 

\bibitem{1}{G.Bertone, D. Hooper, J. Silk, Particle Dark Matter: Evidence, Candidates and Constraints, \em  Phys.Rept.,} {\bf 405} (2005), 279-390.   

\bibitem{2}{G. Bertone, D. Merritt, Dark Matter Dynamics and Indirect Detection, \em Modern Physics Letters A} {\bf 20} (2005), 1021-1036. DOI: 10.1142/S021773205017391

\bibitem{3}{R.Y. Chiao, \em Quantum Theory: A Two-Time Success Story,} Springer, Berlin, 2014.

\bibitem{4}{ D. Clowe, M. Brada$c\check{}$, et al., A direct Empirical Proof of the Existence of Dark Matter, \em The Astrophysical Journal Letters,} {\bf 648} (2006), L109-L113. DOI: 10.1086/508162

\bibitem{5}{L.D.Faddeev, \em Mathematical Aspects of the Three Body Problem in Quantum Scattering Theory,} Davey, New York, 1965.

\bibitem{6}{L.D. Faddeev, S.P. Merkuriev, \em Quantum Scattering Theory for Several Particle Systems,} Kluwer Academic Publishers, Doderecht, 1993.

\bibitem{7}{J.D. Franson, Apparent correction to the speed of light in a gravitational potential, \em New Journal of Physics,} {\bf 16} (2014), 065008, doi:10.1088/1367-2630/16/065008

\bibitem{8}{K. Garrett, G. D$\bar{u}$da, Dark Matter: A Primer, \em Advances in Astronomy,}{\bf  2011} (2011),  ID 968283, pp 22, doi:10.1155/2011/968283 
   
\bibitem{9}{W. Glockle, \em The Quantum Mechanical Few-Body Problem,} Springer, London, 2011.

\bibitem{10}{M.L. Goldberger, K.M. Watson, \em Collision Theory,} Dover Publications, New York, 2004.

\bibitem{11}{S.K. Lee,  et al., Effect of Gravitational Focusing on Annual Modulation in Dark-Matter Direct-Detection Experiments, \em Phys. Rev. Lett.}{\bf 112} (2014), 011301-011305. DOI: 10.1103/PhysRevLett.112.011301

\bibitem{12}{M.L. Mateo, Dwarf Galaxies of the Local Group, \em Annual Review of Astronomy and Astrophysics,} {\bf 36}  (1998), 435-506. DOI: 10.1146/annurev.astro.36.1.435.

\bibitem{13}{C. Moni Bidin, G. Carraro, R. A. Mendez, R. Smith, Kinematical and chemical vertical structure of the Galactic thick disk II. A lack of dark matter in the solar neighborhood,} arXiv:1204.3924v1 [astro-ph.GA]

\bibitem{14}{A.H.G. Peter, Dark Matter,} arXiV:1201.3942v1 [astro-ph.CO], 18 Jan 2012 

\bibitem{15}{A. Pontzen, F. Governato, Cold dark matter heats up, \em Nature, }{\bf 506} (2014), 171 - 178, doi:10.1038/nature12953 

\bibitem{16}{N. Takibayev, Class of Model Problems in Three-Body Quantum Mechanivs That Admit Exact Solutions, \em Physics of Atomic Nuclei,}{\bf 71} (2008), 460. doi:10.1134/S1063778808030083

\bibitem{17}{N. Takibayev, Exact Analytical Solutions in Three-Body Problems and Model of Neutrino Generator, \em EPJ Web of Conferences,} {\bf 3} (2010), doi:10.1051/epjconf/20100305028. 


\end{thebibliography}
\end{document}